\documentclass[prf]{revtex4-2}
\usepackage{tikz}

\begin{document}

\title{Conserved Pseudomomenta in Linear Quasigeostrophic Fluid Flows From Noether's Theorem}
\author{Dušan Beguš}
\affiliation{Department of Physics
Brown University, Providence, Rhode Island 02912-1843, USA}

\author{Chenyu Zhang}
\affiliation{Department of Physics
Brown University, Providence, Rhode Island 02912-1843, USA}

\author{J. B. Marston}
\affiliation{Department of Physics
Brown University, Providence, Rhode Island 02912-1843, USA and 
Brown Center for Theoretical Physics and Innovation, Brown University, Providence, Rhode Island 02912-S, USA}

\date{\today}


\begin{abstract}
Hamiltonian and Lagrangian formulations for the two-dimensional quasi-geostrophic equations linearized about a zonally-symmetric basic flow are presented.  The Lagrangian and Hamiltonian exhibit an infinite U(1) symmetry due to the absence of wave + wave $\rightarrow$ wave interactions in the linearized approximation.  By Noether's theorem the symmetry has a corresponding infinite set of conservation laws which are the well-known pseudomomenta.  There exist separately conserved pseudomomenta at each zonal wavenumber, a point that has sometimes been obscured in past treatments. 
\end{abstract}


\maketitle

\section{Introduction}
\label{intro}
The observation that there are conserved pseudomomenta in the linear approximation to the quasi-geostrophic equation dates back to the works of Arnold \cite{Arnold.1965} and Held and Phillips \cite{Held.1987}. In addition to their practical importance in geophysical fluid dynamics \cite{Mcintyre.1981, Salmon.1998, Bühler_2014}, pseudomomenta also have an intrinsically interesting mathematical feature; in this note we show that the conservation of pseudomomenta in linear wave theory has its origin in the infinite $U(1)$ symmetry of the governing equations. Applying Noether's theorem \cite{Weinberg.1995, Woit.2017} in this context yields pseudomomenta as conserved charges.

Interestingly, the equations of motion of the quasilinear approximation in which the mean-flow responds to Reynolds stresses from the waves also exhibit the same infinite $U(1)$ symmetry as the fully linear equations \cite{Zhang2022,Marston.2023}.  However, we have so far been unable to find Hamiltonian or Lagrangian formulations of the quasilinear approximation, and do not presently know if there are pseudomomenta that continue to be conserved in the quasilinear approximation. If such quantities exist, they would be of great interest as they would have predictive power beyond the linear approximation.  

The organization of this note is as follows.  In Sec. \ref{QG} we review the quasigeostrophic equation and its linearization.  In Sec. \ref{pseudomomenta} we directly show that pseudomomenta of the model presented in Sec. \ref{QG} are conserved by invoking the equations of motion.  In Sec. \ref{Hamiltonian} we present a Hamiltonian approach to the problem \cite{Held.1987}. By defining an appropriate bracket or commutator \cite{Held.1987, Woit.2017}, we show that the pseudomomenta operators commute with the Hamiltonian and are thus conserved.  In Sec. \ref{Lagrangian} we present the Lagrangian for the linear wave theory.  This allows us to construct the pseudomomenta by application of Noether's theorem. The pseudomomenta are the generators of the infinite $U(1)$ symmetry of the Lagrangian. Finally in Sec. \ref{discussion} we discuss the significance of the results.

\section{quasigeostrophic equation of motion and its linearization}
\label{QG}
In the limit of the infinite Rossby deformation length the barotropic quasigeostrophic equation is:
\begin{eqnarray}
\label{EoM}
\partial_{t} \omega = J[\psi, \omega+ f]
\end{eqnarray}
where $\psi$ is the streamfunction of the flow, $\omega = \nabla^2 \psi$ is the vorticity of the flow, and $f= f_0 + \beta y$ is the Coriolis parameter in the $\beta$-plane approximation. $J[\cdot, \cdot]$ is Jacobian operator \cite{Arnold.1965, Woit.2017, Shepherd.1990}:
\begin{eqnarray}
\label{Jacobian}
J[A,B]= \frac{\partial A}{\partial x}\frac{\partial B}{\partial y} - \frac{\partial B}{\partial x}\frac{\partial A}{\partial y}.
\end{eqnarray}
For zonal averages, the standard Reynolds decomposition holds:
\begin{eqnarray}
\label{ReynoldDecomp}
\omega &=& \bar{\omega} + \omega^\prime
\nonumber \\
\psi &=& \bar{\psi} + \psi^\prime 
\label{ReynoldsDecomposition}
\end{eqnarray}
with $\bar{\omega}, \bar{\psi}$ representing the mean vorticity and the streamfunction respectively. Eq. \ref{EoM} then reads:
\begin{eqnarray}
\label{mean_vorticity_EoM}    \partial_{t}\bar{\omega}=\overline{\partial_{t}\omega}= \overline{J[\psi,\omega+ f]}= J[\bar{\psi}, \bar{\omega}]+ \overline{J[\psi^\prime, \omega^\prime]}
\end{eqnarray}
\begin{eqnarray}
\label{vorticity_EoM}
    \partial_{t}\omega^\prime = J[\psi^\prime, \bar{\omega}]+ J[\bar{\psi}, \omega^\prime] + \beta \frac{\partial \psi^\prime}{\partial x} + (J[\psi^\prime, \omega^\prime] - \overline{J[\psi^\prime, \omega^\prime]})\ .
\end{eqnarray}
We now drop the nonlinear last term in Eq. \ref{vorticity_EoM}, $(J[\psi^\prime, \omega^\prime] - \overline{J[\psi^\prime, \omega^\prime]})$ which physically amounts to neglecting wave + wave $\rightarrow$ wave interactions. 
Introducing a Fourier transformation in the zonal direction,
\begin{eqnarray}
\label{vorticity_Fourier}
\omega (x,y,t) &=& \int{\omega(k,y,t) e^{ikx} dk}
\nonumber \\
\psi (x,y,t) &=& \int{\psi(k,y,t) e^{ikx} dk}
\end{eqnarray}
we see that $\bar{\psi}(y, t) =\psi(k = 0, y, t)$, $\bar{\omega}(y, t) = \omega(k = 0,y,t)$ are the zonal averages in the $x$-direction of the streamfunction and vorticity respectively.  (The conventional factors of $2 \pi$ that would appear in the measure are suppressed for clarity.  For a finite spatial domain, one would sum over zonal wavenumber $k$ instead of integrating.)
In the quasi-linear approximation, the last term in Eq. \ref{mean_vorticity_EoM} drives the evolution of the mean flow as follows:
\begin{eqnarray}
\label{quasi_linear_condition}    \partial_{t}\bar{\omega}=\overline{J[\psi^\prime, \omega^\prime]}.
\end{eqnarray}
In the linear approximation we neglect this Reynolds forcing of the mean flow which instead is steady in time. Also, since both $\bar{\psi}, \bar{\omega}$ are $x$-independent, we conclude that $J[\bar{\psi}, \bar{\omega}]=0$ and thus 
\begin{eqnarray}
\label{linear_condition}
    \partial_{t}\bar{\omega}=0.
\end{eqnarray}
The linear equations of motion in $(k, y)$ space are:
\begin{eqnarray}
\label{quasilinear_EoM_vorticity}
\frac{\partial \omega^\prime(k,y,t)}{\partial t} &=& ik \psi^\prime(k,y,t) [\beta +  \frac{\partial \omega(0,y,t)}{\partial y}]
- ik \omega(k,y,t) \frac{\partial \psi(0,y,t)}{\partial y}
\nonumber \\
\frac{\partial \omega^{\prime *}(k,y,t)}{\partial t} &=& -ik \psi^{\prime *}(k,y,t) [\beta + \frac{\partial \omega(0,y,t)}{\partial y}]+ ik \omega^{\prime *} (k,y,t) \frac{\partial \psi(0,y,t)}{\partial y}\ .
\end{eqnarray}

\subsection{Conservation of Pseudomomenta}
\label{pseudomomenta}
Pseudomomenta at zonal wavenumber $k$ are defined to be \cite{Arnold.1965, Held.1987}:
\begin{eqnarray}
\label{pseudomomentum}
    {\cal{M}}(k)= \int{\frac{\omega(k,y,t) \omega^{\prime *}(k,y,t)}{\partial_{y}(\bar{\omega}+f)} ~ dy}\ .
\end{eqnarray}
The integral in Eq. \ref{pseudomomentum} is constant in time for linear quasigeostrophic flows.  Here we directly prove this conservation law by taking the time derivative of Eq. \ref{pseudomomentum} and invoking the equations of motion:
{
\begin{eqnarray}
\label{pseudomomentum_EoM}
\frac{d}{dt}{{\cal{M}}(k)} &=& \int \Big\{\frac{\partial_{t}\omega^\prime(k,y,t) \omega^{\prime *}(k,y,t)+ \omega^\prime(k,y,t) \partial_{t} \omega^{\prime *}(k,y,t)}{\partial_{y}(\bar{\omega} + f)} 
\nonumber \\ 
&-& \frac{\omega(k,y,t) \omega^{\prime *}(k,y,t)}{(\partial_{y}(\bar{\omega}+f))^2}\partial_y \partial_t \bar{\omega} \Big\}~ ~ dy.
\end{eqnarray}}
The second term in the integral vanishes because $\partial_t \bar{\omega}=0$ by Eq. \ref{linear_condition}. So we are left with 
\begin{eqnarray}
\label{pseudomomentum_EoM_simplified}
\frac{d}{dt}{{\cal{M}}(k)} = \int{ \frac{\partial_{t}\omega^\prime(k,y,t) \omega^{\prime *}(k,y,t)+ \omega^\prime(k,y,t) \partial_{t} \omega^{\prime *}(k,y,t)}{\partial_{y}(\bar{\omega} + f)}~ dy}.
\end{eqnarray}
Next, using Eq. \ref{quasilinear_EoM_vorticity}:
\begin{eqnarray}
\label{partial_integration}
\partial_{t}\omega^\prime(k,y,t) \omega^{\prime *}(k,y,t) + \omega^\prime(k,y,t) \partial_{t} \omega^{\prime *}(k,y,t) 
\nonumber \\
= ik [\beta + \partial_{y}\bar{\omega}] \left[\psi^\prime(k,y,t)\omega^{\prime *}(k,y,t)- \psi^{\prime *}(k,y,t) \omega(k,y,t)\right]
\end{eqnarray}
where the $ik \omega(k,y,t) \partial_{y}\bar{\psi} \omega^{\prime *}(k,y,t)$ terms cancel out.
Thus, we end up with the following:
\begin{eqnarray}
\label{pseudomomentum_cancellation1}
\frac{d}{dt}{{\cal{M}}(k)}= \int  ik \frac{\beta + \partial_{y}\bar{\omega}}{\beta + \partial_{y}\bar{\omega}} [\psi^\prime(k,y,t)\omega^{\prime *}(k,y,t) -\psi^{\prime *}(k,y,t) \omega^\prime(k,y,t)] ~ ~ dy.
\end{eqnarray}
Finally, from $\omega = \nabla^2 \psi$ it follows that  $\omega^\prime(k,y,t)= (-k^2 + \partial_y ^2)\psi^\prime(k,y,t)$, and therefore:
{\small
\begin{eqnarray}
\label{pseudomomentum_cancellation2}
\frac{d}{dt}{{\cal{M}}(k)} = \int  ik \frac{\beta + \partial_{y}\bar{\omega}}{\beta + \partial_{y}\bar{\omega}} [\psi^\prime(k,y,t)(-k^2 + \partial_y^2)\psi^{\prime *}(k,y,t)
-\psi^{\prime *}(k,y,t)(-k^2 + \partial_y^2) \psi^\prime(k,y,t)] ~ ~ dy.
\end{eqnarray}}
The terms proportional to $k^2$ cancel each other out, and after a double partial integration of $\psi^\prime(k,y,t)\partial_y^2 \psi^{\prime *}(k,y,t) \to \partial_y^2 \psi^\prime(k,y,t) \psi^{\prime *}(k,y,t)$, we conclude that 
\begin{eqnarray}
\label{pseudomentum_cancellation3}
\frac{d}{dt}{{\cal{M}}(k)} &=& \int ik \frac{\beta + \partial_{y}\bar{\omega}}{\beta + \partial_{y}\bar{\omega}} (-k^2 + \partial_y^2)\psi^\prime(k,y,t) \psi^{\prime *}(k,y,t) ~ dy 
\nonumber \\ 
&-& \int  ik \frac{\beta + \partial_{y}\bar{\omega}}{\beta + \partial_{y}\bar{\omega}} \psi^{\prime *}(k,y,t)(-k^2 + \partial_y^2) \psi^\prime(k,y,t) ~ dy
\nonumber \\
&=& 0\ .
\end{eqnarray}
Thus, ${\cal{M}}(k)$ is conserved in time \cite{Held.1987}.

\section{Hamiltonian Formulation}
\label{Hamiltonian}
We now introduce the Hamiltonian for linear wave theory in Eq. \ref{quasilinear_EoM_vorticity}. The symmetries and conserved quantities of the fully nonlinear Hamiltonian for the quasigeostrophic Eq. \ref{EoM} were studied in \cite{Arnold.1965, Shepherd.1990}. Here we focus on the Hamiltonian for the linear dynamics. It can be inferred from the second-order variation of the energy functional of an inviscid incompressible fluid governed by the barotropic vorticity equation \cite{Arnold.1965}. The Hamiltonian is given by:
\begin{eqnarray}
\label{Hamiltonian_Def}
{H}[\omega^\prime]= \int{\Big\{(\nabla \psi^\prime)^2 + \frac{1}{2}\frac{\partial_{y}\bar{\psi}}{\partial_{y}\bar{\omega}+\beta} (\omega^\prime)^2 \Big\}~ ~ dxdy}\ .
\end{eqnarray}
where $\psi^\prime$ and $\bar{\psi}$ appearing in the integrand satisfy $\nabla^2 \psi = \omega$. With the choice of the Jacobian operator \cite{Arnold.1965, Shepherd.1990, Woit.2017}
\begin{eqnarray}
\label{Jacobian_for_Hamiltonian}
J(\cdot) = \partial(\bar{\omega}+\beta y, \cdot)
\end{eqnarray}
the equation of motion for the waves takes the form
\begin{eqnarray}
\label{Jacobian_EoM}
\partial_{t} \omega^\prime = J\left(\frac{\delta {H}}{\delta \omega^\prime}\right)
\end{eqnarray}
as 
\begin{eqnarray}
\label{EoM_from_Hamiltonian}
\partial_{x}\psi^\prime(\partial_{y}\bar{\omega} + \beta) - \frac{\partial_{y}\bar{\psi}}{\partial_{y}\bar{\omega} + \beta}\partial_{x}\omega^\prime (\partial_{y}\bar{\omega} + \beta) = \partial_{x}\psi^\prime (\partial_{y}\bar{\omega} + \beta)- \partial_{y}\bar{\psi} \partial_{x}\omega^\prime = \partial_{t}\omega^\prime\ ,
\end{eqnarray}
reproducing the linear wave dynamics of Eq.  \ref{quasilinear_EoM_vorticity}. We emphasize that Hamiltonian dynamics  is noncanonical \cite{Shepherd.1990}. Namely, the coordinate $\omega^\prime$ is not accompanied by a canonical momentum; instead, we rely on the symplectic form in Eq. \ref{Jacobian_EoM} to generate the dynamics of the system. 
\subsection{The Infinite U(1) Symmetry of the Hamiltonian}
\label{section_infiniteU1Hamiltonian}
After the Fourier transform in the zonal direction, Eq. \ref{vorticity_Fourier}, the Hamiltonian may be written as a functional of $\omega^\prime(k,y,t)$:
\begin{eqnarray}
\label{Hamiltonian_FourierTransformed}
{H}[\omega^\prime(k,y,t)] = \int  \left( -\psi^{\prime *}(k,y,t)\omega^\prime(k,y,t) + \frac{1}{2}\frac{\partial_{y}\bar{\psi}}{\partial_{y}\bar{\omega}+\beta}\omega^\prime(k,y,t)\omega^{\prime *}(k,y,t)\right) dk dy.
\end{eqnarray}
We notice the following symmetry of the linear Hamiltonian \cite{Marston.2023}: 
\begin{eqnarray}
\label{Hamiltonian_symmetry}
\omega^\prime(k,y,t) \to \omega^\prime(k,y,t)e^{i\theta_k}\ , ~ ~ ~ ~ ~ ~ ~
\omega^{\prime *}(k,y,t) \to \omega^{\prime *}(k,y,t)e^{-i\theta_k}
\end{eqnarray}
which for infinitesimal rotations becomes
\begin{eqnarray}
\label{Hamiltonian_infinitesimal_symmetry}
\omega^\prime(k,y,t) &\to& \omega^\prime(k,y,t)(1+ i \delta\theta_k)
\nonumber \\ 
\omega^{\prime *}(k,y,t) &\to& \omega^{\prime *}(k,y,t)(1-i\delta\theta_k).
\end{eqnarray}
We can write Eq. \ref{Hamiltonian_infinitesimal_symmetry} in the following form:
{\small
\begin{eqnarray}
\label{Hamiltonian_infinitesimalsymmetry_Jacobian}
\omega^\prime(k,y,t) \to \omega^\prime(k,y,t) + i\delta\theta_{-k} \partial \left(\bar{\omega}+ \beta y, \frac{\omega^{\prime *}(-k,y,t)}{\partial_{y}\bar{\omega}+ \beta}\right) =\omega^\prime(k,y,t) + i\delta\theta_{-k} J \left(\frac{\omega^{\prime *}(-k,y,t)}{\partial_{y}\bar{\omega}+ \beta}\right).
\end{eqnarray}}
It is now evident that the pseudomomenta ${\cal{M}}(k)$ 
\begin{eqnarray}
\label{pseudomomentum_Hamiltonian}
{\cal{M}}(k)={\cal{M}}(-k)=\int{\frac{\omega^\prime(k,y,t)\omega^{\prime *}(k,y,t)}{\partial_{y}\bar{\omega}+ \beta} ~dy}
\end{eqnarray}
have the property that 
\begin{eqnarray}
\label{pseudomomentum_functionalDerivative}
\frac{\delta {\cal{M}}(k)}{\delta \omega^\prime (q,y,t)}= \frac{\omega^{\prime *}(k,y,t)}{\partial_{y}\bar{\omega}+ \beta}\delta(k-q)+\frac{\omega^{\prime *}(k,y,t)}{\partial_{y}\bar{\omega}+ \beta}\delta(k+q) 
\end{eqnarray}
which generates the $\frac{\omega^{\prime *}(-q,y,t)}{\partial_{y}\bar{\omega}+ \beta}$ term in Eq. \ref{Hamiltonian_infinitesimalsymmetry_Jacobian} when $q=-k$. So ${\cal{M}}(k)$ generates the infinite U(1) symmetry of Eq. \ref{vorticity_EoM}. Analogous to classical dynamics, we would then expect the pseudomomenta to commute \cite{Shepherd.1990} with the Hamiltonian:
\begin{eqnarray}
\label{bracket_pseudomomentum_Hamiltonian}
    [{\cal{M}}(k), H]=0.
\end{eqnarray}
To establish Eq. \ref{bracket_pseudomomentum_Hamiltonian}, we need to define the notion of a bracket between functionals of $\omega'$.
\subsection{The Vanishing Bracket}
\label{section_vanishing_bracket}
We want to prove Eq. \ref{bracket_pseudomomentum_Hamiltonian} directly. Let us start by motivating the bracket dynamics of Eq. \ref{EoM_from_Hamiltonian}. 
The equation of motion, Eq. \ref{EoM_from_Hamiltonian}, can be expressed as
 \begin{eqnarray}
\frac{\partial}{\partial t}\omega^\prime = [\omega^\prime, H]
\end{eqnarray}
if a bracket between two functionals $A[\omega^\prime]$ and $B[\omega^\prime]$ is defined as:
\begin{eqnarray}
\label{bracket_definition}
[A,B] &=& \int \frac{\delta A}{\delta \omega^\prime(x,y,t)} J \left(\frac{\delta B}{\delta \omega^\prime(x,y,t)}\right) dx dy  
\nonumber \\
&=& \int \frac{\delta A}{\delta \omega^\prime(x,y,t)} \partial \left(\bar{\omega}+ \beta y, \frac{\delta B}{\delta \omega^\prime(x,y,t)}\right) dx dy .
\end{eqnarray}
Fourier transforming along the $x$-axis, Eq. \ref{bracket_definition} becomes:
\begin{eqnarray}
\label{bracket_Fourier}
[A,B] &=& \int \frac{\delta A}{\delta \omega^\prime(k,y,t)} J \left(\frac{\delta B}{\delta \omega^{\prime *}(k,y,t)}\right)dk dy 
\nonumber \\ 
&=& \int  ik \frac{\delta A}{\delta \omega^\prime(k,y,t)}  \frac{\delta B}{\delta \omega^{\prime *}(k,y,t)}  (\partial_{y}\bar{\omega} + \beta)~ dk dy\ .
\end{eqnarray}
Using Eq. \ref{Jacobian_EoM} and Eq. \ref{pseudomomentum_functionalDerivative}:
\begin{eqnarray}
\label{Hamiltonian_derivative}
\frac{\delta {H}}{\delta \omega^{\prime *}(k,y,t)}= -\psi^\prime(k,y,t) + \frac{\partial_{y}\bar{\psi}}{\partial_{y}\bar{\omega}+ \beta}\omega^\prime(k,y,t)
\end{eqnarray}
\begin{eqnarray}
\label{pseudomomentum_Vderivative}
\frac{\delta {\cal{M}}(k)}{\delta \omega^\prime (q,y,t)}= \frac{\omega^{\prime *}(k,y,t)}{\partial_{y}\bar{\omega}+ \beta}\delta(k-q)+\frac{\omega^{\prime *}(k,y,t)}{\partial_{y}\bar{\omega}+ \beta}\delta(k+q)\ . 
\end{eqnarray}
We are now ready to calculate the bracket (Eq. \ref{bracket_pseudomomentum_Hamiltonian}) of the two operators:
\begin{eqnarray}
\label{bracket1}
[{\cal{M}}(k), {H}] &=& \int ik~ \Big\{{\omega^{\prime *}(k,y,t)}\left(-\psi^\prime(k,y,t) + \frac{\partial_{y}\bar{\psi}}{\partial_{y}\bar{\omega}+ \beta}\omega^\prime(k,y,t)\right) 
\nonumber \\ 
&-& {\omega^\prime(k,y,t)}\left(-\psi^{\prime *}(k,y,t) + \frac{\partial_{y}\bar{\psi}}{\partial_{y}\bar{\omega}+\beta} \omega^{\prime *}(k,y,t)\right)\Big\}~dy.
\end{eqnarray}
After some cancelations (mainly of terms of the form $\omega\omega^{\prime *}\frac{\partial_{y}\bar{\psi}}{\partial_{y}\bar{\omega}+ \beta}$), we are left with
\begin{eqnarray}
\label{bracket2}
[{\cal{M}}(k), {H}]= \int{ ik\left[\omega^\prime(k,y,t)\psi^{\prime *}(k,y,t) - \omega^{\prime *}(k,y,t)\psi^\prime(k,y,t)  \right] ~ ~dy}.
\end{eqnarray}
Noting that $(-k^2+ \partial_y^2)\psi = \omega$,
{\small
\begin{eqnarray}
\label{bracket3}
[{\cal{M}}(k), {H}]= \int  ik ~ [(-k^2+ \partial_y^2)\psi^\prime(k,y,t) \psi^{\prime *}(k,y,t) - (-k^2+ \partial_y^2)\psi^{\prime *}(k,y,t) \psi^\prime(k,y,t)] ~ dy
\end{eqnarray}}
which after a partial integration vanishes, confirming that 
\begin{eqnarray}
\label{pseudomomentum_conservation_Hamiltonian}
\frac{d}{dt}{\cal{M}}(k)= [{\cal{M}}(k), {H}]=0\ .
\end{eqnarray}

\section{Lagrangian Formulation}
\label{Lagrangian}
The Lagrangian framework permits the application of Noether's theorem to construct the charges that correspond to symmetries, not just verify the conservation of the charges.  The action that reproduces the equations of motion of the linear theory, Eq. \ref{quasilinear_EoM_vorticity}, is:
\begin{eqnarray}
\label{Lagrangian_Def}
S[\omega^\prime,~ \omega^{\prime *}; \psi^\prime,~ \psi^{\prime *}] &=&
\int~ \Big\{\frac{\omega^{\prime *}(k,y,t) \partial_{t}\omega^\prime(k,y,t)}{\partial_{y}\bar{\omega}+\beta} - ik [\omega^\prime(k,y,t)\psi^{\prime *}(k,y,t) 
+ \omega^{\prime *}(k,y,t)\psi^\prime(k,y,t)] 
\nonumber \\
&+& ik \frac{\partial_{y}\bar{\psi}}{\partial_{y}\bar{\omega}+\beta} \omega^\prime(k, y, t) \omega^{\prime *}(k, y, t)~ +  ik \psi^{\prime *}(k,y,t)\left[(-k^2 + \partial_{y}^2)\psi^\prime(k,y,t)\right] \Big\}~ ~dk dy dt .
\end{eqnarray}
where $\omega^\prime$ and $\psi$ are treated as independent fields. We rely on the equations of motion arising from Eq. \ref{Lagrangian_Def} to generate the relationship $(-k^2+\partial_{y}^2)\psi=\omega$ between the streamfunction and vorticity fields. The action in Eq. \ref{Lagrangian_Def} leads to the following equations of motion, which can be derived using the Euler-Lagrange equations \cite{Weinberg.1995, Woit.2017}:
\begin{eqnarray}
\label{EulerLagrange}
\partial^{i}\left(\frac{\delta S}{\delta (\partial_{i}\omega^\prime(k,y,t))}\right)=\frac{\delta S}{\delta \omega^\prime(k,y,t)}, ~ ~~~\partial^{i}\left(\frac{\delta S}{\delta (\partial_{i}\psi^\prime(k,y,t))}\right)=\frac{\delta S}{\delta \psi^\prime(k,y,t)}
\end{eqnarray}
where the repeated index $i$ is summed over the three space-time coordinates. Note that $\partial_{1}=ik$ due to the Fourier transform in the zonal direction.  For the fields $\omega^{\prime *}, \psi^{\prime *}$ we obtain 
\begin{eqnarray}
\label{EulerLagrange_Fourier}
\partial^{i}\left(\frac{\delta S}{\delta (\partial_{i}\omega^{\prime *}(k,y,t))}\right)=\frac{\delta S}{\delta \omega^{\prime *}(k,y,t)}, ~~~~ \partial^{i}\left(\frac{\delta S}{\delta (\partial_{i}\psi^{\prime *}(k,y,t))}\right)=\frac{\delta S}{\delta \psi^{\prime *}(k,y,t)}
\end{eqnarray}
with $\partial_{1}=-ik$ in this case.
Equations \ref{EulerLagrange} and \ref{EulerLagrange_Fourier} lead to the following equation of motion for the field $\omega^\prime$:
\begin{eqnarray*}
\label{EoM_Lagrangian}
\frac{\partial \omega^\prime(k,y,t)}{\partial t}= ik \psi^\prime(k,y,t) [\beta + \frac{\partial \omega(0,y,t)}{\partial y}] -ik \omega^\prime(k,y,t) \frac{\partial \psi(0,y,t)}{\partial y}
\end{eqnarray*}
which is recognized to be Eq. \ref{quasilinear_EoM_vorticity}, the linear quasigeostrophic wave equation. The equation of motion for $\psi$ reads:
\begin{eqnarray}
\label{streamfunction-vorticity}
(-k^2+\partial_{y}^2)\psi^\prime(k,y,t) = \omega^\prime(k,y,t)
\end{eqnarray}
which is the desired relationship between the streamfunction and vorticity.
The integrand of Eq. \ref{Lagrangian_Def}, which is the Lagrangian, has the $U(1)$ symmetry at each zonal wavenumber $k$:
\begin{eqnarray*}
\label{symmetryLagrangian2}
\omega^\prime(k,y,t) &\to& \omega^\prime(k,y,t) e^{i\theta_k}  
\nonumber \\
\psi^\prime(k,y,t) &\to& \psi^\prime(k,y,t) e^{i\theta_k}
\end{eqnarray*}
which takes the infinitesimal form
\begin{eqnarray}
\omega^\prime(k,y,t) &\to& \omega^\prime(k,y,t) + i\omega^\prime(k,y,t) \delta\theta_{k}
\nonumber \\
\psi^\prime(k,y,t) &\to& \psi^\prime(k,y,t) + i\psi^\prime(k,y,t)\delta\theta_{k}
\end{eqnarray}
Finally, the action in Eq. \ref{Lagrangian_Def} does not have a kinetic term in the field $\psi$. This observation is relevant for the application of Noether's theorem in the next section.
\subsection{Noether's theorem}
\label{section_Noether_Theorem}
Noether's theorem \cite{Weinberg.1995, Woit.2017} states that a continuous symmetry of the action has a corresponding conserved quantity. In our case, for each wavenumber $k$, the conserved charge is \cite{Weinberg.1995}:
\begin{eqnarray}
\label{Noether}
{\cal{M}}(k)= \int{\frac{\delta S}{\delta \left(\partial_{t}\omega^\prime(k,y,t)\right)} \frac{\delta \omega^\prime(k,y,t)}{\delta \theta_{k}}~ dy}.
\end{eqnarray}
Noting that $\delta \omega^\prime(k,y,t)= i \omega^\prime(k,y,t) \delta\theta_k$, along with evaluating $\frac{\delta S}{\delta \left(\partial_{t}\omega^\prime(k,y,t)\right)}= \frac{\omega^{\prime *}(k,y,t)}{\partial_{y}\bar{\omega}+\beta}$, we conclude that there is a family of conserved quantities labeled by wavevectors $k$, colloquially referred to as pseudomomenta \cite{Arnold.1965, Held.1987}:
\begin{eqnarray}
\label{pseudomomentum_Noether}
{\cal{M}}(k)= \int{\frac{\omega^\prime(k,y,t)~ \omega^{\prime *}(k,y,t)}{\partial_{y} \bar{\omega}+ \beta}~ dy}
\end{eqnarray}
with the property that 
\begin{eqnarray}
\label{pseudomomentum_conservation_Noether}
    \frac{d}{dt} {\cal{M}}(k)=0
\end{eqnarray}
We emphasize that the Lagrangian formulation combined with Noether's theorem, Eq. \ref{Noether}, permits the construction of the conserved pseudomomenta. The Hamiltonian approach in Eq. \ref{Hamiltonian_Def} by contrast does not; it only permits one to verify that the already known pseudomomenta are conserved. Hence, the Lagrangian approach offers an advantage in obtaining the conserved pseudomomenta in comparison to the Hamiltonian approach.
\section{Discussion}
\label{discussion}
In this note we presented Hamiltonian and Lagrangian formulations of a barotropic quasigeostrophic fluid in the linear approximation. The existence of a family of conserved quantities, the pseudomomenta, labeled by different zonal wavenumbers was confirmed.  In the literature frequently only the total pseudomomentum, given by a sum over the pseudomomenta at each zonal wavenumber, is considered.  This seems to be because the nonlinear interaction redistributes pseudomomenta between the different zonal wavenumbers, hence the total pseudomomentum can be better conserved than the separate zonal pseudomomenta. However, the conservation is only approximate.

The work reported in this note was motivated by an investigation of the quasilinear approximation in hopes of answering the question of whether or not it too exhibits conserved pseudomomenta since the equations of motion, like those of the linear theory, are invariant under the $U(1)$ rotations. The development of field-theoretic approaches to Eqs. (\ref{vorticity_EoM}) and  (\ref{quasilinear_EoM_vorticity}) were a natural step to addressing the quasi-linear problem. A natural continuation of this work would be to formulate a Lagrangian or Hamiltonian for the quasi-linear approximation, or alternatively to prove that it cannot be done. 

\section{Acknowledgements}
\label{acknowledgements}
We acknowledge helpful discussions with Paul Kushner, Albion Lawrence, Rick Salmon, and Jasper Voss. D.B. acknowledges support from a Karen T. Romer Undergraduate Teaching and Research Award (UTRA).  C. Z. was supported in part by the Simons Foundation (grant no. 62962, GF).  J.B.M. was supported in part by the Simons Foundation (grant nos. 62962, GF and SFI-PD-Pivot Mentor-00008573) and by the US Department of Energy (grant no. DE-SC0024572).

\bibliography{refs}

@article{Shepherd.1990,
  title={Symmetries, {Conservation} {Laws}, and {Hamiltonian} {Structure} in {Geophysical} {Fluid} {Dynamics}},
  author={Theodore G. Shepherd},
  journal={Advances in Geophysics},
  year={1990},
  volume={32},
  pages={287-338},
  url={https://api.semanticscholar.org/CorpusID:118480099}
}

@inproceedings{Arnold.1965,
  title={Conditions for nonlinear stability of stationary plane curvilinear flows of an ideal fluid},
  author={Vladimir I. Arnold},
  year={1965},
  url={https://api.semanticscholar.org/CorpusID:120166862}
}

@article { Held.1987,
      author = "Isaac M.  Held and Peter J.  Phillips",
      title = "Linear and {Nonliear} {Barotropic} {Decay} on the {Sphere}",
      journal = "Journal of Atmospheric Sciences",
      year = "1987",
      publisher = "American Meteorological Society",
      address = "Boston MA, USA",
      volume = "44",
      number = "1",
      doi = "10.1175/1520-0469(1987)044<0200:LANBDO>2.0.CO;2",
      pages=      "200 - 207",
      url = "https://journals.ametsoc.org/view/journals/atsc/44/1/1520-0469_1987_044_0200_lanbdo_2_0_co_2.xml"
}

@article{Marston.2023,
   author = "Marston, J. B. and Tobias, S. M.",
   title = "Recent {Developments} in {Theories} of {Inhomogeneous} and {Anisotropic} {Turbulence}", 
   journal= "Annual Review of Fluid Mechanics",
   year = "2023",
   volume = "55",
   number = "Volume 55, 2023",
   pages = "351-375",
   doi = "https://doi.org/10.1146/annurev-fluid-120720-031006",
   url = "https://www.annualreviews.org/content/journals/10.1146/annurev-fluid-120720-031006",
   publisher = "Annual Reviews",
   issn = "1545-4479",
   type = "Journal Article",
  }

@article{Mcintyre.1981, 
title={On the {‘wave} {momentum’} {myth}}, 
volume={106}, DOI={10.1017/S0022112081001626}, 
journal={Journal of Fluid Mechanics}, 
author={Mcintyre, M. E.}, 
year={1981}, 
pages={331–347}}

@book{Salmon.1998,
  title={Lectures on {Geophysical} {Fluid} {Dynamics}},
  author={Salmon, R.},
  isbn={9780195108088},
  lccn={97001811},
  url={https://books.google.com/books?id=yozmCwAAQBAJ},
  year={1998},
  publisher={Oxford University Press, Oxford, UK}
}

@book{Weinberg.1995, 
place={Cambridge}, 
title={The {Quantum} {Theory} of {Fields}}, 
publisher={Cambridge University Press, Cambridge, UK}, 
author={Weinberg, Steven}, 
year={1995}}

@book{Woit.2017,
  title={Quantum {Theory}, {Groups} and {Representations}: {An} {Introduction}},
  author={Woit, P.},
  isbn={9783319646121},
  url={https://books.google.com/books?id=G248DwAAQBAJ},
  year={2017},
  publisher={Springer International Publishing, New York, USA}
}

@phdthesis{Zhang2022,
  author = {Zhang, Chenyu},
  title = {Simulation of {Turbulence} in {Quasilinear Models}},
  school = {Brown University Department of Physics, Providence, Rhode Island},
  year = {2022},
  type = {{PhD} {Thesis}}
}

@book{Bühler_2014, place={Cambridge}, edition={2}, series={Cambridge Monographs on Mechanics}, title={Waves and Mean Flows}, publisher={Cambridge University Press, Cambridge, UK}, author={Bühler, Oliver}, year={2014}, collection={Cambridge Monographs on Mechanics}}

\end{document}